\documentclass{moriond}

\usepackage{amsmath}
\usepackage{hyperref}
\usepackage{epsfig,bbm} 

\usepackage[bf, sf, FIGTOPCAP, nooneline, tight]{subfigure}

\def\healpix {{\tt HEALPix}~}
\def\nside {N_{\rm nside}}

\newcommand{\figref}[1]{Figure. (\ref{#1})}
\newcommand{\Photo}{}

\begin{document}
\vspace*{4cm}

\author{Yabebal T. Fantaye}
\address{Departement of Mathematics, University of Tor Vergata Roma2, Rome, Italy
Email: \textit{fantaye@mat.uniroma2.it}}
\address{Part of the talk is presented on behalf of the Planck Collaboration.}
\title{Test of cosmic isotropy in the Planck era}

\maketitle           

\abstracts{
The two fundamental assumptions in cosmology are that the Universe is
statistically homogeneous and isotropic when averaged on large
scales. Given the big implication of these assumptions, there has been
a lot of statistical tests carried out to verify their
validity. Since the first high-precision Cosmic Microwave Background
(CMB) data release by the WMAP satellite, many anomalies that
challenges the isotropy assumption, including dipolar power asymmetry
on large angular scales, have been reported.  In this talk I will
present a brief summary of the test of cosmic isotropy we carried out
in the latest WMAP and Planck temperature data.
}

\section{Introduction}

 

The European Space Agency (ESA) Planck satellite mission has recently
produced the most accurate picture of the Universe by measuring the
CMB with unprecedented precision. The scientific findings of this
mission is presented in a series of science papers, which are mostly
consistent with that of WMAP, the previous CMB satellite experiment by
NASA. These
papers \footnote{\href{http://sci.esa.int/planck/51551-simple-but-challenging-the-universe-according-to-planck/.}{http://sci.esa.int/planck/51551-simple-but-challenging-the-universe-according-to-planck/.}}
present a simple but challenging picture of the Universe. Despite
being consistent with the standard picture, the Universe seen by
Planck has some anomalies whose interpretation might require a new
physics.

Amongst the Planck confirmed CMB anomalies the hemispherical power
asymmetry \cite{eriksen2004,hansen2004} is one of the major one.
This anomaly implies that the distribution of power in one side of the
universe is different from that of the opposite one, leading to a
breakdown of cosmic isotropy. In general, it is assumed that on large
scales, scales that are not processed by non-linear gravity, the
Universe is homogeneous and isotropic. The former implies that if we
are able to zoom out and look at larger patches of the Universe, the
statistical property of these patches should be the same; while the
latter means if we set ourselves at one point, like on earth, and see
all around us the statistical property of the universe in one
direction should be similar to another one, hence rotation invariant.

The significance of the hemispherical power asymmetry, however, has
often been called into question, in particular, due to the alleged
\emph{a-posteriori} nature of the statistics used; the significance of
the anomaly is calculated using a statistical method that is designed
to detect the observed feature. The challenge to this criticism was
first given by \cite{hansen2009} who showed that the 5-year WMAP data show
a similar trend up to $\ell=600$.  Moreover, using an alternative
approach which modelled the power asymmetry in terms of a dipolar
modulation field, as suggested by \cite{gordon:2005},
\cite{hoftuft2009} found a $3.3\sigma$ detection using data smoothed
to an angular resolution of $4.5^{\circ}$ FWHM, with an axis in
excellent agreement with previous results.

In this talk I will present a brief summary of the test of cosmic
isotropy we carried out in the latest WMAP and Planck temperature
data. 



\section{Power asymmetry analysis methods}

There are in general two methods employed to test power asymmetry in
the CMB map. The first one is to compute local-power on a disc at
different sky directions and compare their consistency with isotropic
power distribution. The effect of mask, noise and other complications
are incorporated in such a method by using a set of simulations which
are processed in a similar way to the data. The excess mismatch
between the distribution of local-power in the data and simulations
serve as a measure of significance of the power asymmetry. The second
one is to assume a dipolar or higher order power asymmetry model and
do a likelihood analysis. The two common analyses in this category are
the pure dipole modulation by \cite{hoftuft2009} and the generalised
dipole modulation model as implemented in Bipolar spherical harmonics
(BiPoSHs) technique \cite{biposh2006}.

In our analyses of the WMAP and Planck data in
\cite{magnus2013,varasym}, we quantified the degree
of anisotropy by measuring the distribution of local-power, which is
measured as the variance or power spectrum of the local patches, in
the CMB maps. The variance of a map is related to its power spectrum as

\begin{equation}
\sigma^2 = \frac{1}{4\pi}\sum_{\ell=0}^{\ell_{max}}(2\ell+1)C_\ell,
\end{equation}

and hence using both of these quantities offers a possibility to study any deviation from isotropy
in both real and harmonic space. Moreover,
using power spectrum allows us easily study the scale dependence of a
possible power variation in the CMB maps, while using variance we
avoid mask induced complications in harmonic decomposition.

To incorporate a scale dependence study to the variance based method,
we decomposed the map in to needlet components.
\begin{equation}\label{eqn:needfield}
\beta _{j}(n)=\sum_{\ell =B^{j-1}}^{B^{j+1}}b^{2}(\frac{\ell }{B^{j}})T_{\ell }(n),   
\end{equation}


where $T_{\ell }(n)$ denotes the component at multipole $\ell $ of the CMB
map $T(n)$, e.g.%
\[
T(n)=\sum_{\ell }T_{\ell }(n) ,  
\]%
$n\in S^{2}$ denotes the pointing direction, $B$ is a fixed parameter
(usually taken to be between $1$ and $2$) and $b(.)$ is a smooth function
such that $\sum_{j}b^{2}(\frac{\ell }{B^{j}})=1$ for all $\ell .$

The advantage of using needlets is that the needlet filter has a very good localisation
property both in pixel and harmonic space, and the needlet component maps are minimally affected
by masked regions, especially at high-frequency. In particular, of
course the multipole components $T_{\ell }(n)$ cannot be reconstructed on
masked data; nevertheless their linear combination \eqref{eqn:needfield} can be
shown to be very robust to the presence of gaps, and more so on
small scales/high frequencies.

In what follows we will refer the local-power method with variance
measure as \emph{local-variance} and local-power with power spectra
measures as \emph{local-$C_{\ell}$}. A summary of the procedures we
followed in both of these methods are as follows:

\begin{enumerate}

\item create a binary patch mask which are centred on low resolution
  \healpix \cite{Healpix2005} pixels. For both local-$C_\ell$ and local-variance methods
  we have considered 3072 ($\nside=16$) highly overlapping disc
  patches covering the entire sky. The radius of the discs varies from
  1 to 90 degrees. The final local-$C_\ell$ results, however, are
  quoted from analysis performed using 12 non-overlapping
  patches, the base \healpix $\nside=1$ pixels. We have shown that there is no significant
  difference between the results obtained using overlapping or
  non-overlapping patches.

\item looping over patch numbers, create an analysis patch mask by
  multiplying the binary disc mask with that of a foreground confidence
  and point source masks. In the case of local-$C_\ell$ analysis, we
  have apodised the foreground confidence and point source masks to reduce
  correlations between modes.

\item apply the analysis patch mask to the CMB + noise maps which are
  either the WMAP and Planck data or realistic simulations.

\item for each patch compute $C_\ell$s using the {\tt MASTER}
  technique \cite{hivon2002} in bins of 16 multipoles; or variance of
  the unmasked pixels.  For the local-$C_\ell$ case, the {\tt MASTER}
  $16-\ell$ blocks are further binned into $100-\ell$ blocks to
  reduce bin to bin correlations as well as to compare our results with 
  previous similar analyses.
  

\item for each patch estimate the mean and variance of local- $C_\ell$
  or local-variance using $N_{sim}$ realistic simulations. The mean is
  used to subtract the local mean-field from both data and
  simulations, while the variance is used to weight the corresponding patch.

\item estimate dipole amplitude and dipole directions of the
  local-power map using the \healpix routine {\tt remove\_dipole} by
  applying an inverse variance weighting.

\end{enumerate}

\section{Results and Discussion}

In this section I will present the summary of results presented in
\cite{magnus2013,varasym}, and some updates which are
not included in these published papers. In particular, I will describe
the result we obtained using the local-variance analysis on needlet decomposed maps.  

As outlined in the previous section, our test statistics is based on
dipole amplitudes and dipole directions of the data and simulation
local-power maps. For local-$C_\ell$s, which are estimated in
100-$\ell$ blocks, we obtained $\ell_{\rm max}/100$ $\nside=1$
local-$C_\ell$ maps. For WMAP and Planck $\ell_{\rm max}=600$
and $\ell_{\rm max}=1500$ is used, which are corresponding to 6 and 15 dipole amplitude
and directions, respectively. On the other hand, for the
local-variance analysis we used 20 disc sizes with radius ranging from
1-90 degrees, and obtained 20 dipole amplitudes and directions. The
later case is estimated for both the original real-space maps and the
corresponding needlet decomposed maps.

In all cases we
have used p-values to quantify the significance of the observed power
asymmetry.  The p-values are computed by comparing the anisotropic
signal measured by our test-statistics, magnitude of dipole amplitudes
or clustering of dipole directions, with those obtained from realistic
simulations. For the WMAP case, we generated 1000 CMB-plus-noise Monte
Carlo (MC) simulations based on the WMAP9 best-fit $\Lambda$CDM power
spectrum \cite{wmap9_2}. Noise realisations are drawn as uncorrelated
Gaussian realisations with a spatially varying RMS distribution given
by the number of observations per pixel. For Planck we adopt the 1000
``Full Focal Plane'' (FFP6) end-to-end simulations produced by the
Planck collaboration based on the instrument performance and noise
properties. These simulations also incorporate lensing and component
separation effects. The simulations are treated identically to the
data in all steps discussed below.

Using dipole modulated simulations, we found that clustering of dipole
directions is more sensitive to measuring a small power asymmetry
signal, small dipole modulation amplitudes, than comparing the
magnitude of dipole amplitudes in data and simulations. The dipole
directions clustering, however, can only be obtained when there are at
least a few number of independent dipole direction estimates, which is
only the case for local-$C_\ell$ analysis.  
 For the local-variance method, however, the dipole directions for different
disc sizes are correlated, and hence we can not use the alignment of
dipole directions as a measure of significance. We compute
p-values, therefore, based on excess of the data dipole amplitude from those of
the simulations.


\subsection{local-$C_\ell$ analysis}
In \cite{magnus2013} we performed a local-$C_\ell$ analysis on WMAP
9-year data using 12 patches. We found that the power asymmetry is
statistically significant at the $3.2\sigma$ confidence level for
$\ell=2$--600, where the data is signal dominated. The preferred
asymmetry direction is $(l,b)=(227,-27)$, consistent with previous
claims. Individual asymmetry axes estimated from six independent
multipole ranges are all consistent with this direction.

In the same paper, we did also MCMC analysis on the 12 independent
patches to determine the local best-fit values of the six $\Lambda$CDM
cosmological parameters: the Hubble constant today ($H_0$); fractional
densities of baryons ($\Omega_b$), cold dark matter ($\Omega_m$) and
cosmic curvature ($\Omega_k$); the reionization optical depth
($\tau$); the amplitude ($A_s$) and spectral index ($n_s$) of the
initial scalar fluctuations. Since most of the information used to
constrain these parameters comes from the CMB power spectrum, the
observed power asymmetry may have been caused by asymmetry in one or
more of these cosmological parameters. After doing local-MCMC
analysis, we obtained a map of local best-fit values for each six
parameters. Since we used all multipole blocks of the {\tt MASTER}
power spectra for a given patch to estimate the parameters, we obtain
only a single dipole amplitude and direction per parameter per map. As
a result of this we can not use the dipole directions to measure
significance of parameter asymmetry. The significance of parameter
asymmetry is, therefore, measured by comparing the data and simulation
local-best-fit map dipole amplitudes. We found that none of the
cosmological parameters show a significant asymmetry. This is probably
because of the large error bars in the parameters.  It is, however,
interesting to note that the dipole direction for the parameters
$A_s$, $n_s$ and $\Omega_b$ are in alignment with the power asymmetry
direction, implying these parameters are the most sensitive to the
power asymmetry observed.

\begin{figure}[t]
\begin{center}
\includegraphics[width=1\textwidth,clip=true]{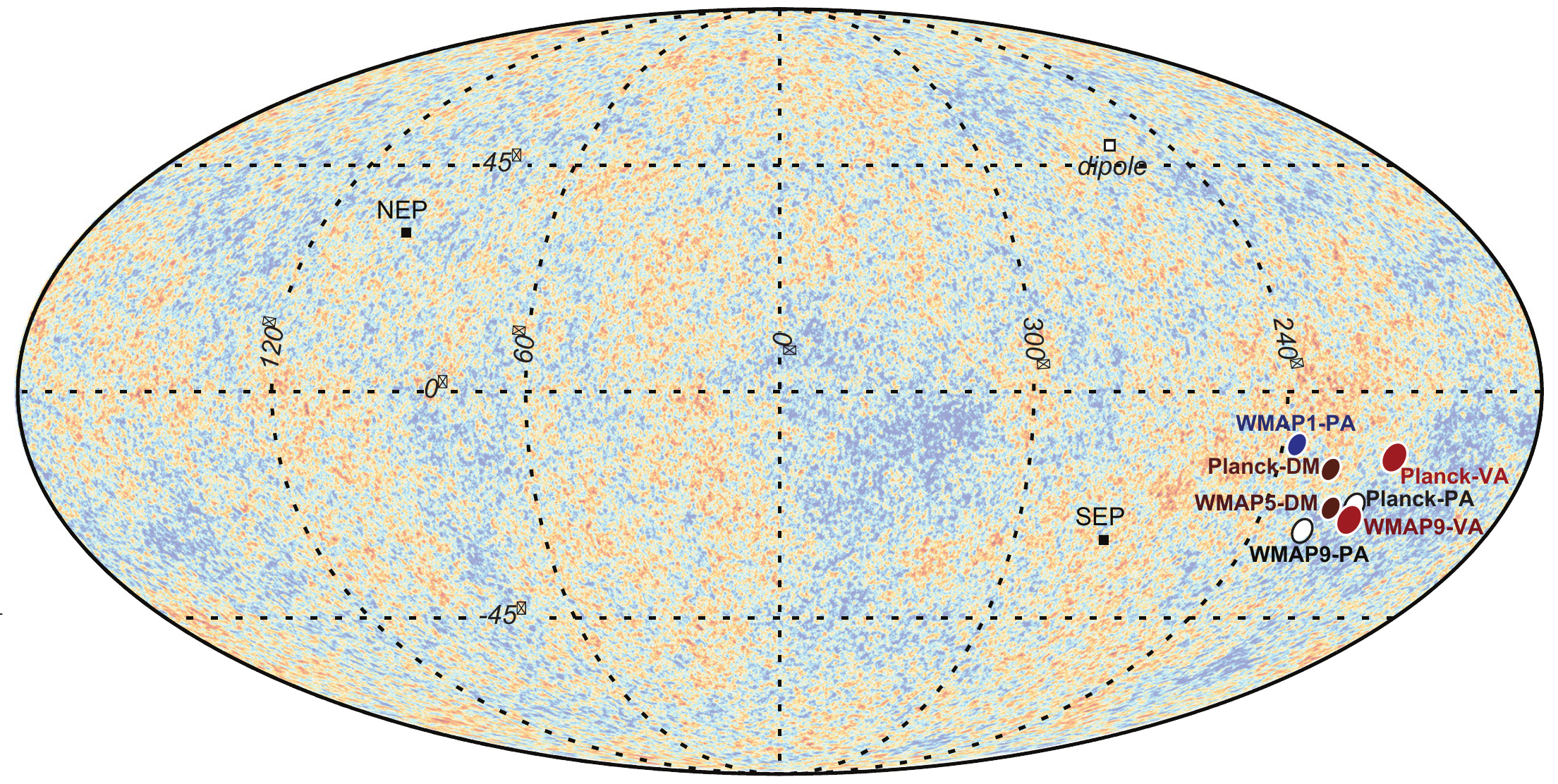} \\
\caption{\footnotesize{Asymmetry directions found in different
    hemispherical power asymmetry analysis. The local variance of the
    WMAP 9-year and Planck 2013 data [denoted by WMAP9-VA and
      Planck-VA], as well as the directions found previously from the
    latest likelihood analyses of the dipole modulation model [denoted
      by WMAP5-DM and Planck-DM], and the local-power spectrum analyses
    [denoted by WMAP1-PA, WMAP9-PA and Planck-PA] for the
    WMAP and Planck data. The background map is the CMB sky observed
    by Planck (SMICA). VA, DM and PA stand for variance asymmetry,
    dipole modulation and power asymmetry, respectively. This figure
    is taken from Akrami et. al. 2013$^9$.} \label{fig:alldirections}} 
\end{center}
\end{figure}

The local-$C_\ell$ analysis of the Planck collaboration \cite{planck2013_IandS}
confirms a similar power asymmetry observed in the WMAP data,
reassuring the power asymmetry is not due to systematic effects. This
agreement across a wide range of scales as well as two different data
sets clearly removes the statistical \emph{a-posterior} interpretation
of the effect, and poses a new challenge for a detail investigation of
the effect. In the Planck paper it was shown that the power asymmetry extends up to $\ell \sim 600$ with a
significance of at least $3 \sigma$. Beyond this scale, however, the doppler
boosting, which is due to the motion of our solar system barycenter
with respect to the CMB, becomes dominant. Boosting is a significant contamination for power asymmetry study,
and it has to be taken into account when looking for power asymmetry
at small scales, large multipoles.

\subsection{local-variance analysis}
\begin{figure}[t]
\begin{center}
\includegraphics[width=0.45\linewidth,clip=true]{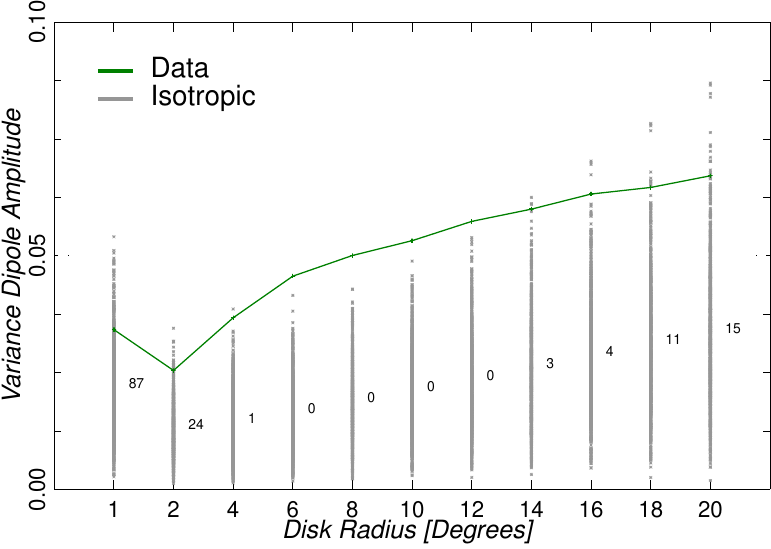}
\includegraphics[width=0.45\linewidth, clip=true]{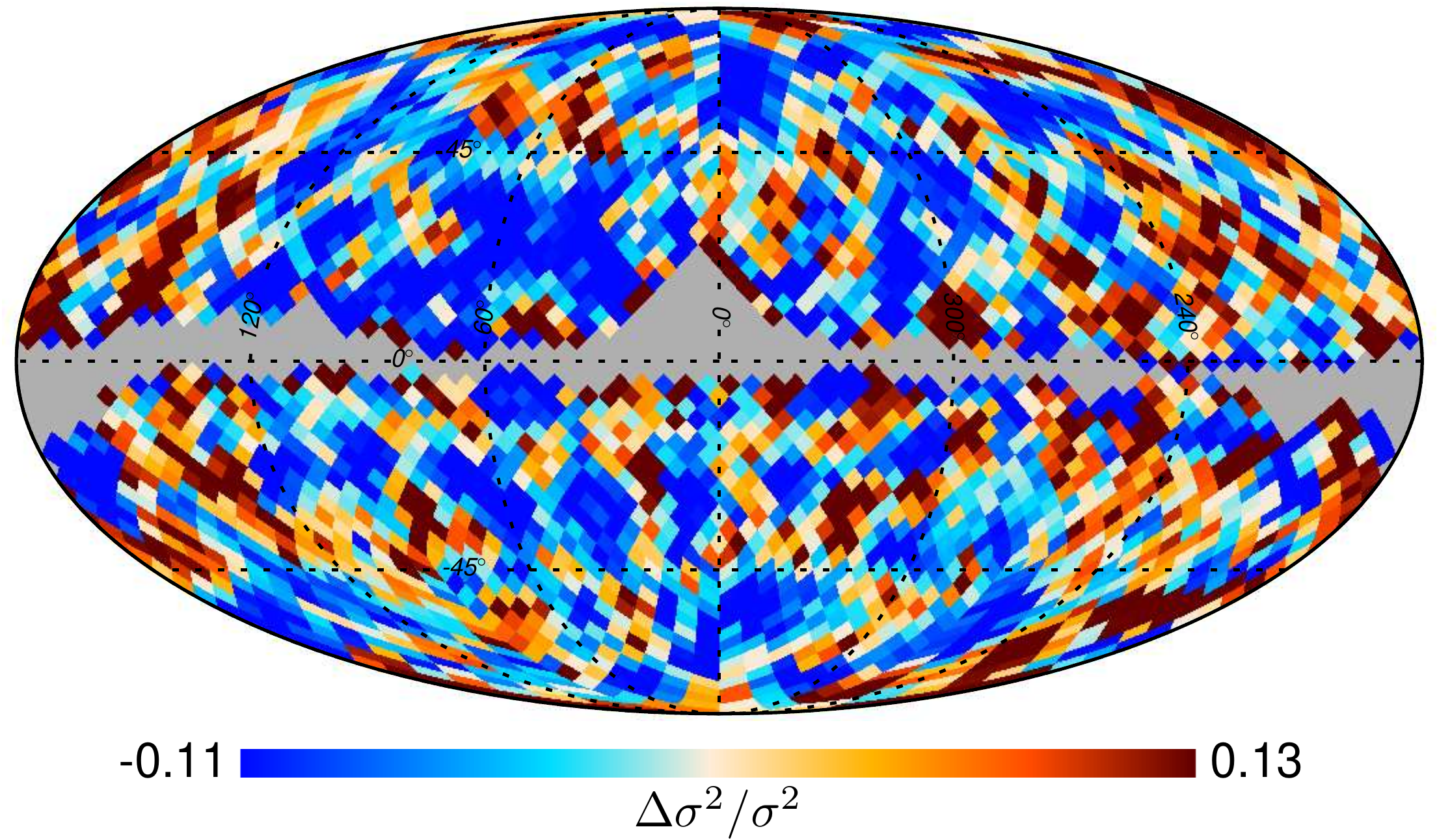}
\caption{Left panel: local-variance dipole amplitude as
    a function of disc radius for  Planck {\tt SMICA}
    data (in green) versus the 1000 isotropic FFP6 simulations (in
    grey). The labels above each scale indicate the number of
    simulations with amplitudes larger than the ones estimated from
    the data, and are located at the means of the amplitude values
    from the simulations. Right panel: mean-field subtracted,
    local-variance map computed with $6^\circ$ discs for Planck
    data. This figure is taken from Akrami et. al. 2013$^9$. \label{fig:varplanck}} 
\end{center}
\end{figure}

In \cite{varasym} we performed a local-variance analysis in both WMAP
9-year and Planck 1-year data.  Left panel of \figref{fig:varplanck} shows
the results for Planck, which compares the dipole
amplitudes of the data local-variance map (green stars) with that of the 1000 FFP6 
simulations (grey stars) - none of the 1000 isotropic simulations have
local-variance dipole amplitudes larger than the data over the range 
$6^\circ \le r_{\small {\textrm{disc}}}\le 12^\circ$. This implies that the
variance in the Planck data exhibits a dipolar- like spatial
variations that are statistically significant (at least) at the 
$\sim 3-3.5\sigma$ level or a $p$-value of less than 0.001. We
showed that the distribution of the simulation dipole amplitudes
are well fit by a Gaussian distribution for all discs sizes.  
The right panel of \figref{fig:varplanck} visually illustrates
mean-field subtracted local-variance map for the $6^\circ$ radius
disc. The dipole direction obtained from this map is
$(l,b)=(212^\circ,-13^\circ)$.

Similar analysis using the WMAP 9-year data yields a statistical
significance of $\sim 2.9\sigma$, and a direction fully consistent
with those derived from Planck and other previous analyses. Our check
shows that the difference in mask and input power spectra used to simulate
the WMAP simulations seems to drive the difference in significance
between Planck and WMAP, and a further investigation of this issue is necessary 
once the slight tension between WMAP and Planck data is
resolved.

We have checked that the local-variance based results do not
significantly change when we account for the doppler boosting effect
either by including boosting in the simulations or by deboosting the
data, similar to what we did for the Planck 143GHz channel map with
the local-$C_\ell$ analysis.  This is mostly because the
local-variance method is mostly dominated by large scale modes where
the effect of doppler boosting is small. As we will show below, however, this
method actually is able to detect the boosting signal with $>3 \sigma$
significance when considering a high-pass filtered map.

The preferred directions obtained using local-variance and
local-$C_\ell$ methods from our papers as well as a number of similar
results obtained in previous papers are summarised in Figure
\ref{fig:alldirections}.

\begin{figure} 
\begin{center}
\includegraphics[width=0.8\linewidth, trim = 0 -10 00 0, clip=true]{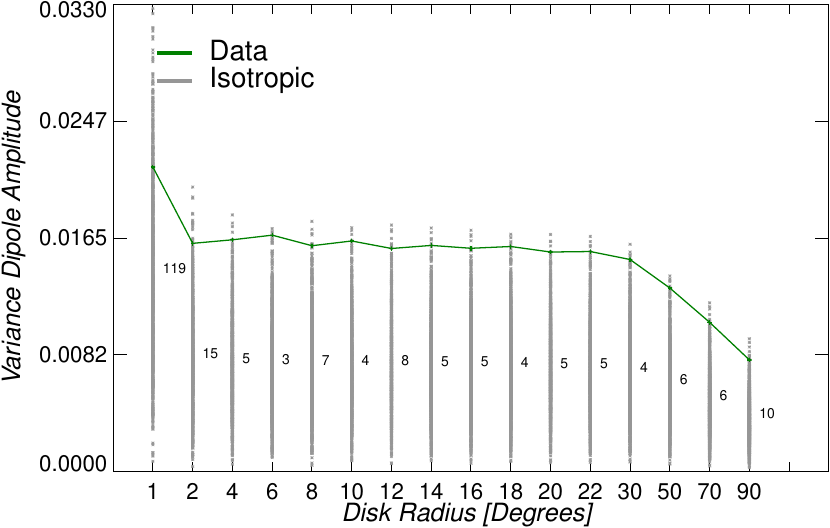}
\caption{Dipole amplitudes of the local-variance map for the $j=9$, $256 \leq \ell \leq 1024$, 
    needlet component  as a function of disc radius. The Planck ({\tt
      SMICA}) data (in green) versus the 1000 isotropic FFP6
    simulations (in grey). The labels above each scale indicate the
    number of simulations with amplitudes larger than the ones
    estimated from the data. \label{fig:needlet} }
\end{center}
\end{figure}

\begin{figure} 
\begin{center}
\includegraphics[width=0.9\linewidth, trim = 0 -10 0 0, clip=true]{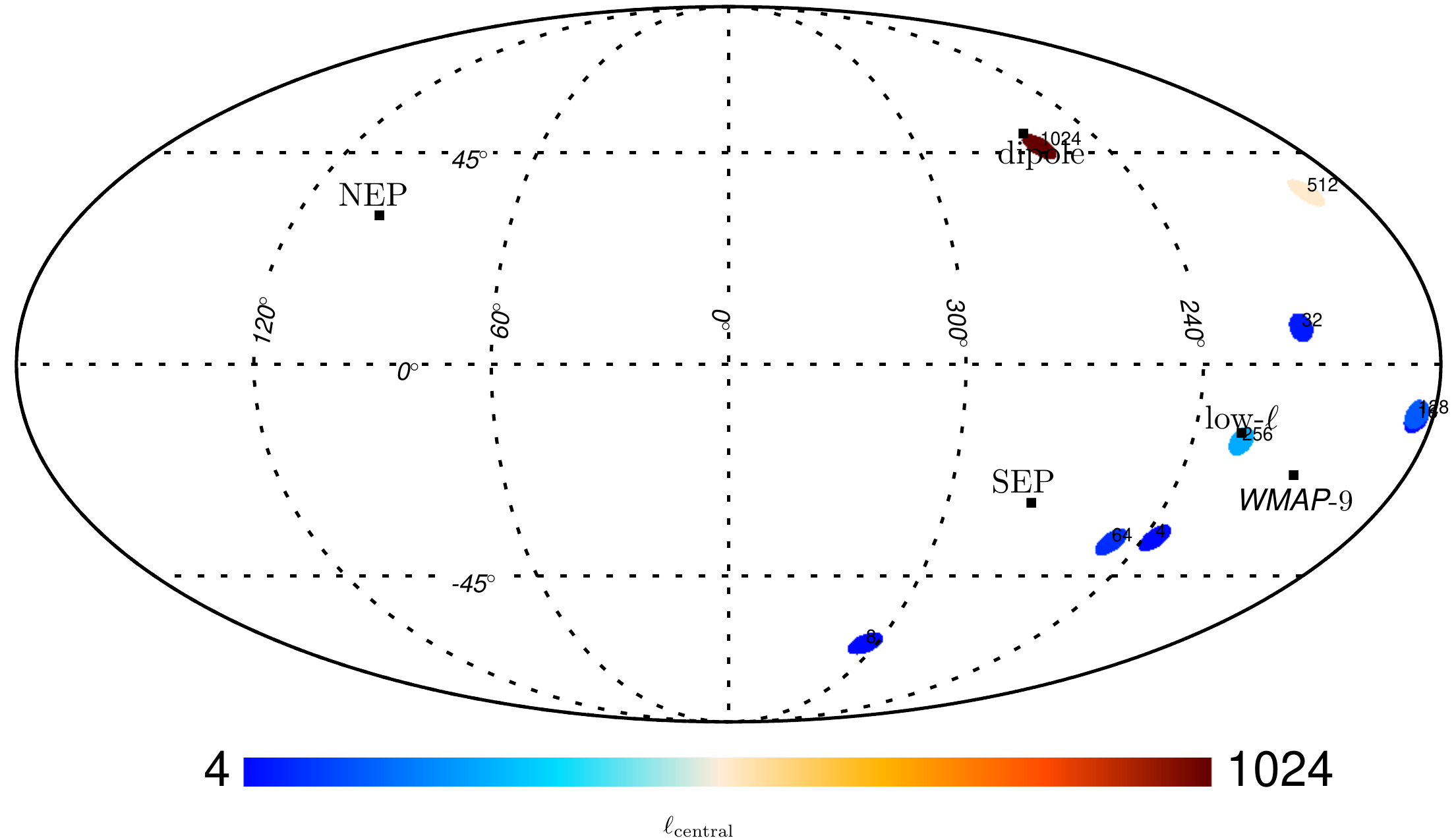}
\caption{Dipole
    directions of  the local-variance maps of the nine needlet components, $j=2,3,\ldots, 10$ with a disc radius of $90\circ$. The
    labels in this plot indicates the central multipole of a given needlet
    map.  \label{fig:needletB} }
\end{center}
\end{figure}

\subsection{Local-variance analysis on needlet components}

To study the scale dependence of power asymmetry using the
local-variance method, we decomposed the Planck SMICA data and FFP6 simulations
into needlet components using \eqref{eqn:needfield}, and performed the
above analysis on each of the decomposed maps.
The needlet parameters we used are such that a given map 
is decomposed into nine needlet component maps each having a
compact support over a multipole range defined by $\ell = [B^{j-1},
  B^{j+1}]$ where $j=2,3,..10$ and $B=2$.

As shown in \figref{fig:needlet}, all but the $j=9$ needlet component
have dipole amplitudes consistent with the FFP6 simulations - hence no
significant power asymmetry. The $j=9$ needlet map, which has a support to
multipole ranges $\ell = [256,1024]$, shows a significant
anisotropy with p-value of 0.003, and with a dipole direction close to
the CMB dipole. This means what we have detected in this component is
not the power asymmetry we talked about in the previous paragraphs,
but the well-know doppler boosting effect caused by our motion with
respect to the CMB frame. This effect has been detected for the first
time by Planck using a completely different algorithm
\cite{planck1_boosting}, and is used to measure our solar system
barycenter velocity independent of the CMB dipole component. In
addition, as shown in  \figref{fig:needletB}, the
dipole direction of the $j=10$ component has a dipole direction right
on the CMB dipole. The Planck data local-variance dipole amplitude for
this particular needlet component, however, is compatible with the FFP6
simulations. The reason for this may be due to the fact that the
majority of the multipoles covered by the $j=10$ component are
sensitive to the boosting signal but with a small signal to noise
values, while all of multipoles in the $j=9$ component are signal
dominated but with less sensitivity to the boosting signal, or
possibly contaminated by a non-vanishing power asymmetry signal.

Some interesting features to note from
\figref{fig:needlet} \& (\ref{fig:needletB}) are: the dipole
directions for the needlet component $j=2-8$ are close to the power
asymmetry direction; the dispersion of dipole amplitudes for isotropic
simulations are large for small radius discs and becomes small for
large discs. This is opposite to what we saw when using the full
map. Of course, this is understandable since most of the needlet maps
are not affected by the cosmic variance of the large scale modes.

The implication of Doppler boosting detection by our method implies
that at least our method is sensitive to dipolar power asymmetry with
amplitudes up to $0.1\%$. It is also a strong validation to our entire
pipeline. Moreover, clustering of the lower multipole dipole
directions, needlet components $j=2-8$, to the power asymmetry dipole
direction may hint existence of asymmetry at intermediate
multipoles. Of course, a lot of verification work has to be done in
this regard.


\section{Summary}
In this talk I have discussed the unique nature of the CMB in testing
the statistical property of the Universe. Thanks to the high precision
experiments like Planck, the test of cosmological principles, the
assumption of statistical isotropy and homogeneity, is now becoming a
major field in the CMB and large scale data analysis.

The current CMB data hints a lopsided Universe, more large
scale structures in one side of the Universe than the other. The
significance of this anomaly, however, is low, $< 4\sigma$, so one
can not rule out yet the effect being just a statistical
fluke. Moreover, although the persistence of the anomaly in both WMAP
and Planck data seems to suggest the cause is unlikely to be
foreground or systematic effects, we can not yet fully exclude the
possibility of a local Universe phenomena. For this and other cases
more work needs to be done to confirm the signal we are observing is
truly cosmological.  If that is proven, then this observation will
represent another major discovery about the nature of our Universe and
might lead to a new insight about the inner workings of inflation,
which is responsible for laying out the initial conditions of the
Universe from a mere quantum fluctuations.

In this endeavour there are different models in the literature trying
to explain what is observed in the data, but to date there is no any
single theoretical model that explains all the observations, power
asymmetry both at large and intermediate scales. 

While theorists are working out a viable model, there are going to be
results from different experiments. For example, the full Planck
temperature and polarisation data will be released in near future and
it will be interesting to see what the outcomes will be. The planned
CMB experiment PRISM \cite{prism} promises to deliver a high-precision
CMB polarisation maps. Since the statistical nature of polarisation
maps as well as its systematics and foregrounds are not necessarily
similar to that of the temperature maps, PRISIM will be crucial in
testing the isotropy hypothesis. On the other hand, EUCLID
\cite{euclid}, a space-based survey mission from ESA, will map the
large scale structure with an unprecedented precision. Such
observations will be able to test the isotropic assumption in the
large scale structure, hence providing an independent
verification. The future possibilities of the study and
characterisation of the statistical nature of our Universe is
therefore bright. These studies will ultimately contribute to our
understanding of the processes that shaped the Universe from its birth
to where it is now and where it is going.

\section*{Acknowledgments}
This conference proceeding is a summary of the work I did in collaboration with M. Axelsson,  
Y. Akrami, A. Shafieloo, H. K. Eriksen, F. K. Hansen, A. J. Banday and K. M. G\'{o}rski. 
This work is supported by ERC Grant 277742 Pascal. I acknowledge the use of resources from the
Norwegian national super-computing facilities, NOTUR. Maps and results
have been derived using the \healpix (http://healpix.jpl.nasa.gov)
software package developed by \cite{Healpix2005}.

\newpage

\section*{References}

\bibliography{yabebal_cmbanomaly}

\end{document}